\input harvmac
\input epsf
\def\inbar{\,\vrule height1.5ex width.4pt depth0pt}
\def\IC{\relax\hbox{$\inbar\kern-.3em{\rm C}$}}
\def\IR{\relax{\rm I\kern-.18em R}}
\def\IP{\relax{\rm I\kern-.18em P}}

\newcount\figno
\figno=0
\def\fig#1#2#3{
\par\begingroup\parindent=0pt\leftskip=1cm\rightskip=1cm\parindent=0pt
\baselineskip=11pt
\global\advance\figno by 1
\midinsert
\epsfxsize=#3
\centerline{\epsfbox{#2}}
\vskip 12pt
{\bf Fig.\ \the\figno: } #1\par
\endinsert\endgroup\par
}
\def\figlabel#1{\xdef#1{\the\figno}}
\def\encadremath#1{\vbox{\hrule\hbox{\vrule\kern8pt\vbox{\kern8pt
\hbox{$\displaystyle #1$}\kern8pt}
\kern8pt\vrule}\hrule}}

\noblackbox
\Title{\vbox{\baselineskip12pt\hbox{hep-th/0205209}
\hbox{SU-ITP-02/15,~~SLAC-PUB-9201} }}
 {\vbox{ {\centerline{Bouncing Brane Cosmologies }
\smallskip
\centerline{from Warped String
Compactifications}} } }

\centerline{Shamit Kachru$^{1}$ and Liam McAllister$^{2}$ }
\medskip
\centerline{$^{1}$Department of Physics and SLAC}
\centerline{Stanford University} \centerline{Stanford, CA 94305/94309}
\centerline{Email:~skachru@stanford.edu}
\smallskip
\centerline{$^{2}$ Department of Physics}
\centerline{Stanford University} \centerline{Stanford, CA 94305}
\centerline{Email:~lpm@stanford.edu}
\medskip
\noindent We study the cosmology induced on a brane
probing a warped throat region in a Calabi-Yau compactification of
type IIB string theory. For the case of a BPS D3-brane probing the
Klebanov-Strassler warped deformed conifold, the cosmology described
by a suitable brane observer is a bouncing, spatially
flat Friedmann-Robertson-Walker universe with time-varying Newton's
constant,  
which passes smoothly from a contracting to an expanding phase. In the
Klebanov-Tseytlin
approximation to the Klebanov-Strassler solution the cosmology would
end with a big crunch singularity. In this sense, the warped
deformed conifold provides a string theory resolution of a
spacelike singularity in the brane cosmology. The four-dimensional
effective action appropriate for a brane observer is a simple
scalar-tensor theory of gravity.  In this description of the physics,
a bounce is possible because the relevant energy-momentum
tensor can classically violate the null energy condition.

\Date{May 2002}

\lref\hawking{S.W. Hawking and G.F.R. Ellis, {\it The~Large~Scale
Structure~of~Spacetime}, Cambridge University Press, 1973.}

\lref\KK{A. Kehagias and E. Kiritsis, ``Mirage Cosmology,'' JHEP {\bf
9911}
022 (1999), hep-th/9910174.}
\lref\RS{L. Randall and R. Sundrum, ``A Large Mass Hierarchy from a
Small Extra Dimension,'' Phys. Rev. Lett. {\bf 93} 3370 (1999),
hep-th/9905221.}
\lref\veneziano{M. Gasperini and G. Veneziano, ``Pre-Big Bang in
String Cosmology,'' Astropart. Phys. {\bf 1} 317 (1993),
hep-th/9211021.}

\lref\KO{C. Herzog, I. Klebanov and P. Ouyang, ``Remarks on the
Warped Deformed Conifold,'' hep-th/0108101.}

\lref\mukher{S. Mukherji and M. Peloso, ``Bouncing and Cyclic Universes
from Brane Models,'' hep-th/0205180.}

\lref\fakir{R. Fakir, ``General Relativistic Cosmology with No Beginning
of Time,'' gr-qc/9810054.}

\lref\kallosh{K. Dasgupta, C. Herdeiro, S. Hirano and R. Kallosh,
``D3/D7 Inflationary Model and M-theory,'' hep-th/0203019.}

\lref\ekpy{J. Khoury, B. Ovrut, P. Steinhardt and N. Turok,
``The Ekpyrotic Universe: Colliding Branes and the Origin of
the Hot Big Bang,'' Phys. Rev. {\bf D64} 123522 (2001),
hep-th/0103239\semi
R. Kallosh, L. Kofman and A. Linde, ``Pyrotechnic Universe,''
Phys. Rev. {\bf D64} 123523 (2001), hep-th/0104073.}

\lref\lust{D. L\"ust, ``Cosmological String Backgrounds,'' hep-th/9303175\semi
C. Kounnas and D. L\"ust, ``Cosmological String Backgrounds from Gauged WZW
Models,'' Phys. Lett. {\bf B289}, 56 (1992), hep-th/9205046.} 

\lref\brustein{R. Brustein and R. Madden, ``Graceful Exit and Energy Conditions
in String Cosmology,'' Phys. Lett. {\bf B410}, 110 (1997), hep-th/9702043.}

\lref\youm{D. Youm, ``Closed Universe in Mirage Cosmology,''
Phys. Rev. {\bf D63}, 085010 (2001), hep-th/0011290.}

\lref\typez{E. Papantonopoulos and I. Pappa, ``Type 0 Brane Inflation
from Mirage Cosmology,'' Mod. Phys. Lett. {\bf A15}, 2145 (2000),
hep-th/0001183\semi
P. Brax and D.A. Steer, ``Non-BPS Brane Cosmology,'' hep-th/0204120.}

\lref\CDO{P. Candelas and X. de la Ossa, ``Comments on Conifolds,''
Nucl. Phys. {\bf B342}, 246 (1990).}

\lref\anti{S. Kachru, J. Pearson and H. Verlinde, ``Brane/Flux
Annihilation and the String Dual of a Nonsupersymmetric Field
Theory,'' hep-th/0112197.}

\lref\audschaf{J. Audretsch and G. Sch\"afer, ``Thermal Particle
Production in a Contracting and Expanding Universe Without Singularity,''
Phys. Lett. {\bf A66} 459 (1978).}

\lref\GKP{S.B. Giddings, S. Kachru and J. Polchinski, ``Hierarchies from
Fluxes in String Compactifications,'' hep-th/0105097.}
\lref\KS{I. Klebanov and M. Strassler, ``Supergravity and a Confining
Gauge Theory: Duality Cascades and $\chi$SB resolution of Naked
Singularities,'' JHEP {\bf 0008} 052 (2000), hep-th/0007191.}
\lref\Herman{H. Verlinde, ``Holography and Compactification,''
Nucl. Phys. {\bf B580} 264 (2000), hep-th/9906182\semi
C. Chan, P. Paul and H. Verlinde, ``A Note on Warped String
Compactification,'' Nucl. Phys. {\bf B581} 156 (2000), hep-th/0003236.}
\lref\Others{
K. Dasgupta, G. Rajesh and S. Sethi, ``M-theory, Orientifolds and
G-flux,'' JHEP {\bf 9908} 023 (1999), hep-th/9908088\semi
B. Greene, K. Schalm and G. Shiu, ``Warped Compactifications in
M and F theory,'' Nucl. Phys. {\bf B584} 480 (2000), hep-th/0004103\semi
P. Mayr, ``Stringy World Branes and Exponential Hierarchies,''
JHEP {\bf 0011} 013 (2000), hep-th/0006204.}

\lref\evaetal{O. Aharony, M. Fabinger, G. Horowitz and E. Silverstein,
``Clean Time-dependent String Backgrounds from Bubble Baths,''
hep-th/0204158.}
\lref\Bell{Bellucci and Faraoni}
\lref\Mend{Mendes and Mazumdar, Brans-Dicke Brane Cosmology}
\lref\bekenstein{J. Bekenstein, ``Exact Solutions of Einstein-Conformal
Scalar Equations,'' Ann. Phys. {\bf 82} 535 (1974)\semi
J. Bekenstein, ``Nonsingular General-relativistic Cosmologies,'' Phys.
Rev.
{\bf D11} 2072 (1975).}
\lref\NK{Kaloper, Singularities in Scalar-Tensor Cosmologies }
\lref\Quiros{Quiros, Bonal, Cardenas, Brans-Dicke-type theories and
avoidance of the cosmological singularity }
\lref\jatkar{For example see: S. Bhattacharya, D. Choudhury, D. Jatkar and A.A. Sen,
``Brane Dynamics in the Randall-Sundrum Model, Inflation and Graceful
Exit,''
hep-th/0103248.}
\lref\VB{M. Visser and C. Barcelo, ``Energy Conditions and their
Cosmological Implications,'' gr-qc/0001099.}
\lref\Paris{C. Molina-Paris and M. Visser, ``Minimal Conditions for the
Creation of a Friedmann-Robertson-Walker Universe from a `Bounce',"
Phys. Lett. {\bf B455} 90 (1999), gr-qc/9810023.}
\lref\nek{N. Nekrasov, ``Milne Universe, Tachyons, and Quantum Group,''
hep-th/0203112.}

\lref\lms{H. Liu, G. Moore and N. Seiberg, ``Strings in a Time-dependent
Orbifold,'' hep-th/0204168.}
\lref\kut{S. Elitzur, A. Giveon, D. Kutasov and E. Rabinovici,
``From Big Bang to Big Crunch and Beyond,'' hep-th/0204189.}

\lref\KT{I.R. Klebanov and A.A. Tseytlin, ``Gravity Duals of
Supersymmetric $SU(N) \times SU(N+M)$ Gauge Theories,'' Nucl. Phys.
{\bf B578} 123 (2000), hep-th/0002159.}
\lref\seiberg{N. Seiberg, ``From Big Crunch to Big Bang: Is it
Possible?,''
hep-th/0201039\semi
J. Khoury, B. Ovrut, N. Seiberg, P. Steinhardt and N. Turok, ``From
Big Crunch to Big Bang,'' Phys. Rev. {\bf D65} 086007 (2002),
hep-th/0108187.}
\lref\vafa{R. Brandenberger and C. Vafa, ``Superstrings in the
Early Universe,'' Nucl. Phys. {\bf B316} 391 (1989).}
\lref\alexds{A. Buchel, ``Gauge/gravity Correspondence in an
Accelerating Universe,'' hep-th/0203041.}
\lref\KW{I. Klebanov and E. Witten, ``Superconformal Field Theory
on Three-branes at a Calabi-Yau Singularity,'' Nucl. Phys.
{\bf B536} 199 (1998), hep-th/9807080.}

\lref\Gubser{S. Gubser, ``AdS/CFT and Gravity,'' Phys. Rev.
{\bf D63} 084017 (2001), hep-th/9912001.}
\lref\Randall{C. Csaki, M. Graesser, L. Randall and J. Terning,
``Cosmology of Brane Models with Radion Stabilization,''
Phys. Rev. {\bf D62} 045015 (2000), hep-th/9911406.}

\lref\globular{S.~Degl'Innocenti, G.~Fiorentini, G.G.~Raffelt, B.~Ricci
and A.~Weiss,
``Time variation of Newton's constant and the age of globular
clusters,''
Astron. Astrophys. {\bf 312} 345 (1996),
astro-ph/9509090.}

\lref\binetruy{ P. Bin\'etruy, C. Deffayet, U. Ellwanger and D. Langlois,
``Brane cosmological evolution in a bulk with cosmological constant,''
Phys. Lett. {\bf B477} 285 (2000), hep-th/9910219.}
\lref\mukohyama{ S. Mukohyama, ``Brane-world solutions, standard
cosmology, and dark radiation,''
  Phys. Lett. {\bf B473} 241 (2000), hep-th/9911165.}
\lref\khoury{ J. Khoury and R.-J. Zhang, ``On the Friedmann Equation in
Brane-World Scenarios,'' hep-th/0203274.}

\lref\medved{ A.J.M. Medved, ``Big Bangs and Bounces on the Brane,''
hep-th/0205037.}

\lref\grinstein{B. Grinstein, D.R. Nolte and W. Skiba, ``On a Covariant
Determination of Mass Scales in
 Warped Backgrounds,'' hep-th/0012074.}
\lref\cornalba{L. Cornalba, M.S. Costa and C. Kounnas, ``A Resolution of
the Cosmological Singularity with Orientifolds,''
  hep-th/0204261.}

\lref\costa{L. Cornalba and M. Costa, ``A New Cosmological Scenario
in String Theory,'' hep-th/0203031.}

\lref\vijay{V. Balasubramanian, S. Hassan, E. Keski-Vakkuri
and A. Naqvi, ``A Space-Time Orbifold: A Toy Model for a Cosmological
Singularity,'' hep-th/0202187.}

\lref\bcraps{B. Craps, D. Kutasov and G. Rajesh, ``String Propagation in 
the Presence
of Cosmological Singularities,'' hep-th/0205101.}

\newsec{Introduction}

There has recently been considerable interest in the properties of 
string theory cosmology.  A generic feature of general
relativistic cosmologies is the presence of singularities, which is 
guaranteed under a wide range of circumstances by the
singularity theorems \hawking.
Since string theory has had great success in providing physically
sensible descriptions of certain timelike singularities in
compactification geometries, one can hope that it will similarly
provide insight into the spacelike or null singularities which
arise in various cosmologies. Proposals
in this direction have
appeared
in e.g. 
\refs{\vafa,\veneziano,\lust,\brustein,\seiberg,\vijay,\nek,\costa,\lms,\kut,\cornalba,\bcraps}.

In a slightly different direction, the possibility of localizing
models of particle physics on three-branes in a 
higher-dimensional
bulk geometry
has motivated a great deal of work on brane-world cosmology
(see
\refs{\KK,\Gubser,\Randall,\ekpy,\kallosh} and
references therein for various examples).
Of particular interest to us will be the ``mirage'' cosmology \KK\ which
is experienced by a D3-brane observer as he falls through a bulk
string theory background.  In this note,
we present a simple and concrete example where such an observer
would describe a cosmology which evades the singularity theorems:
his universe is a flat FRW model which smoothly interpolates between
a collapsing phase and an expanding phase.

The background through which the D3-brane moves is a Klebanov-Strassler
(KS) throat region \KS\ of a IIB Calabi-Yau compactification.
Compactifications including such throats, described in \GKP, yield
models with 4d
gravity and a warp factor which can vary by many orders of
magnitude as one moves in the
internal space (as in the proposal of Randall and Sundrum (RS) \RS).
The backgrounds discussed in \GKP\ would also admit, in many cases,
some number of wandering D3-branes.
Such a brane can fall down the KS throat and bounce smoothly back out,
as the supergravity background has small curvature everywhere.  
The induced
cosmology on this probe, as described by an observer who holds
particle masses ${\it fixed}$, is a spatially flat 
Friedmann-Robertson-Walker
universe which
begins in a contracting phase, passes smoothly through a minimum scale
factor, and then re-expands.\foot{A different approach to using the
KS model to generate an interesting string theory cosmology recently
appeared in
\alexds.}
A D3-brane probe in this background satisfies a ``no-force'' condition
which makes it possible to control the velocity of
the contraction; in addition, the background can be chosen so that the
universe is large in Planck units at the bounce.  For this reason, the
calculations which lead the brane observer to see a bounce are
controlled and do not suffer from large stringy or quantum gravity
corrections.
It is important to note that in this scenario, the effective 4d Newton's 
constant $G_N$ varies with the scale factor of the universe; 
this results from the varying overlap of the graviton
wavefunction with the D3-brane.

The KS solution is actually a stringy resolution of the singular
Klebanov-Tseytlin (KT) supergravity solution \KT, which ends with
a naked singularity in the infrared.  A brane falling into a
Klebanov-Tseytlin throat would therefore undergo a singular big
crunch. In this sense, the cosmology we study involves a stringy
resolution of a spacelike singularity, from the point of view of an
observer on the brane.

Although one can describe the cosmological history of these 
universes using the behavior of the induced metric along the brane
trajectory, it is also interesting to consider the 4d effective
field theory that a brane resident could use to explain his 
cosmology. We construct a simple toy model
of these cosmologies using a
4d 
scalar-tensor theory of gravity. The
scalar can be identified with the open string
scalar field $\Phi_{r}$ (corresponding to radial motion down the
warped throat) in the Born-Infeld action for the D3-brane.
It is well known that such scalar-tensor theories can
classically violate the null energy condition,
making a bounce possible. Related facts about scalar field theories
coupled to gravity have been exploited previously by Bekenstein
and several
subsequent authors \refs{\bekenstein,\Paris,\fakir,\VB}. 

The organization of this note is as follows.  In \S2\ we use 
the construction of \GKP\ to study the cosmology on a brane
sliding down the KS throat.
In \S3\ we provide a discussion of
the effective scalar-tensor theory of gravity a brane theorist would
probably
use to explain his observations.
We close with some thoughts on further directions in \S4.

Several previous authors have investigated the possibility of bounce
cosmologies in scalar-tensor theories and in brane-world models. 
For FRW models with spherical spatial sections ($k=+1$), 
examples in various contexts have appeared in \refs{\bekenstein,\Paris,
\fakir}.  As we were completing this paper, other
discussions of bounces
in brane-world models 
appeared in \refs{\medved,\mukher}.  To the best of 
our knowledge, this note
provides the first controlled example in string theory of a 
bouncing,
spatially flat FRW cosmology with 4d gravity. 

\newsec{Brane cosmology in a warped Calabi-Yau compactification}

\subsec{The compactifications}

In \refs{\Herman,\GKP,\Others}, warped
string compactifications were explored as a means
of realizing the scenario of Randall and Sundrum \RS\ in a string
theory context.  It was shown that
compactifications of IIB string theory on Calabi-Yau orientifolds
provide the necessary ingredients.  In such models, one derives a
tadpole condition of the form \eqn\tadpole{ {1\over 4} N_{O3}
~=~N_{D3} +
{1\over {2(2\pi)^4 (\alpha^\prime)^2}}\int_{X} H_{3}\wedge F_{3}~.}
Here $X$ is the
Calabi-Yau manifold, $N_{O3}$ and $N_{D3}$ count the number of
orientifold planes coming from fixed points of the orientifold
action and the number of transverse D3-branes, and $H_{3},F_{3}$
are the NSNS and RR three-form field strengths of the IIB
theory.\foot{In an F-theory description, the left-hand side of
\tadpole\ is replaced by ${\chi(X_{4})\over 24}$, where $X_4$ is
the relevant elliptic Calabi-Yau fourfold.} In general, the
left-hand side of \tadpole\ is nonzero and can be a reasonably
large number, giving rise to the possibility of compactifications
with large numbers of transverse D3-branes or internal flux
quanta. Since both of these lead to nontrivial warping of the
metric as a function of the internal coordinates, \tadpole\ tells
us that these Calabi-Yau orientifolds provide a robust setting for
finding warped string compactifications \refs{\Herman,\GKP,\Others}.

We can make this somewhat vague statement much more precise in the example
of
the warped deformed conifold.  The conifold geometry is defined in
${\IC}^{4}$ by
\eqn\singcon{{z_{1}}^2+{z_{2}}^2+{z_{3}}^2+{z_{4}}^2 = 0.}
It is topologically a cone over $S^2 \times S^3$; we will refer to the
direction transverse to the base as the ``radial
direction'' (with small $r$ being close to the tip and large $r$ being
far out along the cone).
The $\it deformed$ conifold geometry
\eqn\conifold{{z_{1}}^2+{z_{2}}^2+{z_{3}}^2+{z_{4}}^2 = \epsilon^2}
has two nontrivial 3-cycles, the $A$-cycle $S^3$ which collapses as
$\epsilon \to 0$, and the dual $B$-cycle.
Klebanov and Strassler found that the infrared region of the geometry
which is holographically dual to a
cascading $SU(N+M) \times SU(N)$ ${\cal N}=1$ supersymmetric gauge
theory is precisely a warped version of the deformed conifold geometry,
with
nontrivial 3-form fluxes
\eqn\flux{{1\over {(2\pi)^2 \alpha^\prime}}\int_{A} F = M, ~~
{1\over {(2\pi)^2 \alpha^\prime}}\int_{B} H = - k}
and $N=kM$.
In particular, the space \conifold\ is non-singular and the smooth geometry
dual to the IR of the gauge theory reflects the confinement of the
Yang-Mills theory (with the small parameter $\epsilon$ mapping to
the exponentially small dynamical scale of the gauge theory).  In a
cruder approximation to the physics,
Klebanov and Tseytlin
had earlier found a dual gravity description with a naked
singularity \KT; this heuristically corresponds to the unresolved
singularity in \singcon.

In \GKP, the warped, deformed conifold with flux
\conifold, \flux\ was embedded in string/F-theory compactifications to
4d.
The small $r$ region is as in \KS, while at some large $r$ (in the UV
of the dual cascading field theory), the solution is glued into a
Calabi-Yau manifold.
The fluxes give rise to a potential which fixes (many of)
the Calabi-Yau moduli
(and in particular the $\epsilon$ in \conifold), while the fluxes plus in
some
cases wandering D3-branes saturate the tadpole condition \tadpole.
If one considers one of the cases with $N_{D3} > 0$, then it is natural
to imagine a cosmology arising on a wandering D3-brane as it falls down
towards the tip of the conifold \conifold.

\subsec{The Klebanov-Strassler geometry}

The KS metric is given by (we use the conventions of \KO)
\eqn\ksmetric{
ds^2 = h^{-1/2}(\tau) \eta_{\mu\nu}dx^{\mu} dx^{\nu} +
h^{1/2}(\tau)ds_{6}^{2}}
where $ds_{6}^{2}$ is the metric of the deformed conifold,
\eqn\wdefcon{ds_{6}^{2}={1\over 2}\epsilon^{4/3} K(\tau) ( {1\over 3
K^{3}(\tau)} [ 
d\tau^2 + (g^5)^2 ] + {\rm cosh}^2 ({\tau\over 2}) [(g^3)^2 + (g^4)^2]+
{\rm sinh}^2 ({\tau\over 2})[(g^1)^2 + (g^2)^2]).}
Here
\eqn\ggdef{
\eqalign{
g^1 = {{e^1 - e^3}\over \sqrt{2}},~~ & g^2 = {{e^2 - e^4}\over
\sqrt{2}}\cr
g^3 = {{e^1 + e^3}\over \sqrt{2}},~~ & g^4 = {{e^2 + e^4}\over
\sqrt{2}}\cr
& g^5 = e^5}}
where
\eqn\eedef{
\eqalign{
& e^1 = -sin(\theta_1) d\phi_1,~~e^2 = d\theta_1\cr
& e^3 = cos(\psi) sin(\theta_2) d\phi_2 - sin(\psi) d\theta_2\cr
& e^4 = sin(\psi)sin(\theta_2) d\phi_2 + cos(\psi) d\theta_2\cr
& e^5 = d\psi + cos(\theta_1) d\phi_1 + cos(\theta_2) d\phi_2~.}}
$\psi$ is an angular coordinate which ranges from $0$ to $4\pi$,
while $(\theta_1,\phi_1)$ and $(\theta_2,\phi_2)$ are the
conventional coordinates
on two $S^2$s.
The function $K(\tau)$ in \ksmetric\ is given by
\eqn\kdef{K(\tau) = {({\rm sinh}(2\tau) - 2\tau)^{1/3} \over {2^{1/3}
{\rm sinh}(\tau)}}~.}
Clearly in \ksmetric\ $\tau$ plays the role of the ``radial'' variable
in the conifold geometry, with large $\tau$ corresponding to large $r$.

Finally, the function $h(\tau)$ in \ksmetric\ is rather complicated; it
is given by the expression
\eqn\his{h(\tau) ~=~(g_s M \alpha^\prime)^{2} 2^{2/3}
\epsilon^{-8/3}I(\tau)}
where
\eqn\iis{I(\tau) ~=~\int_{\tau}^{\infty} dx
{{x {\rm coth}(x) - 1}\over {{\rm sinh}^{2}(x)}
} ({\rm sinh}(2x) - 2x)^{1/3}.}
It will be useful to note that this reaches a maximum at $\tau = 0$ and
decreases monotonically as $\tau \rightarrow \infty$.
There are also
nontrivial backgrounds of the NSNS 2-form and RR 2-form potential; their
detailed form will not enter here, but they are crucial in understanding
why the D3-brane propagates with no force in the background \ksmetric.

Since the form of $h(\tau)$ will be important in what follows, we take
a moment here to give some limits of the behavior of
formulae \his,\iis \KO.  For very small $\tau$, one finds $I(\tau) \sim
a_0 + O(\tau^2)$, with $a_0$ a constant of order 1.  In this limit the
complicated metric \ksmetric\
simplifies greatly (c.f. equation(67) of \KO):
\eqn\kstwo{
\eqalign{
ds^2 \to {\epsilon^{4/3}\over {2^{1/3} a_{0}^{1/2} g_s M \alpha^\prime}}
dx_n dx_n + & a_{0}^{1/2} 6^{-1/3} (g_s M \alpha^\prime) ({1\over
2}d\tau^2
+ {1\over 2}(g^5)^2 + (g^3)^2 + (g^4)^2 \cr
& + {1\over 4}\tau^2 [(g^1)^2
+ (g^2)^2])~.}} This is $R^{3,1}$ times (the small $\tau$ limit of) the
deformed conifold.  In
particular, the $S^{3}$ has fixed radius proportional to $\sqrt{g_sM}$,
and so
the curvature can be made arbitrarily small for large $g_sM$.
In the opposite limit of large
$\tau$, the metric simplies to Klebanov-Tseytlin
form.  Introducing the coordinate $r$ via
\eqn\rdef{r^2 = {3\over 2^{5/3}} \epsilon^{4/3} e^{2\tau \over 3}} and
using the asymptotic behavior $I(\tau) \sim 3 \times 2^{-1/3}
(\tau - {1\over 4}) e^{-{4\tau \over 3}}$,
one finds
\eqn\otherlim{ds^2 \to
{r^2\over{L^2 \sqrt{ln(r/r_s)}}} dx_n dx_n + {L^2 \sqrt{ln(r/r_s)}\over
r^2} dr^2 + L^2 \sqrt{ln(r/r_s)} ds_{T^{1,1}}^2}
where
$ds_{T^{1,1}}^2$ is the metric on the Einstein manifold $T^{1,1}$
and $L^2 = {9g_s M \alpha^\prime \over {2\sqrt{2}}}$.
 This means
that up to logarithmic corrections, the large $\tau$ behavior
gives rise to an $AdS_5$ metric for the $x^{\mu}$ and $\tau$ directions.
This is the expected behavior from the field theory dual, since large
$\tau$ corresponds to the UV, where the theory is approximately the
Klebanov-Witten ${\cal N}=1$ SCFT \KW.

\subsec{Trajectory of a falling brane}

We will start the D3-brane at some fixed $\tau = \tau^*$
and send it flying towards
$\tau=0$ with a small initial proper velocity $v$ in the radial $\tau$
direction.  Before describing the trajectory we will briefly explain our
notation.  $\tau$ always indicates the
radial coordinate in the KS geometry \ksmetric\ and is dimensionless in
our conventions.
We will reserve $t$ for proper time (for the infalling brane)
and $\dot{}$ for $d\over{dt}$, while $\xi$ represents the coordinate time,
in terms of which the metric is
\eqn\ximetric{ds^{2}=h(\tau)^{-{1\over{2}}}(-{d\xi}^2 + \sum_i
dx_i^{2})+g_{\tau\tau}{d\tau}^2+angles}
and thus
\eqn\propertime{{ {({{dt}\over{d\xi}})}^{2} }=h(\tau)^{-{1\over{2}}}
{(1-h(\tau)^{{1\over{2}}}g_{\tau\tau} {{({ {d\tau}\over{d\xi}
})}^{2}})}~.}
To leading order in the velocity we have ${({{dt}\over{d\xi}})}^{2}\approx 
h(\tau)^{-{1\over{2}}}$.

Proper distance is given by $d = \int{d\tau' g_{\tau\tau}^{1/2}}$,
and proper velocity by $v\equiv \dot{d}=\dot{\tau}g_{\tau\tau}^{1/2}$.
The initial values of the position, proper distance, coordinate velocity,
and proper velocity
are denoted by $\tau_{*},d_{*},\dot{\tau}_{0}$ and $v_0$, respectively.

The
D3-brane trajectory is determined by the Born-Infeld action
\eqn\borninfeld{S_{BI}={-1\over{g_s^{2} l_s^{4} }} \int {d^{3}\sigma d\xi
~\left( h (\tau)^{-1}
\sqrt{
1-h(\tau)^{{1\over{2}}} g_{\tau\tau} {({ {d\tau}\over{d\xi} })} ^{2} }
-h(\tau)^{-1}\right)}}
where we have neglected contributions from the U(1) gauge field on the
brane.
At leading order in a low-velocity expansion, rewritten in terms of
derivatives with respect to proper time,
\eqn\dbi{S_{BI}={1\over{2 g_s^{2}
l_s^{4}}}\int{d^{3}\sigma d\xi ~h(\tau)^{-1}g_{\tau\tau}{\dot{\tau}^{2}}}}
where the cancellation of the potential $h(\tau)^{-1}$ is the realization
of the no-force condition.
Conservation of energy then yields
\eqn\taudot{{\dot{\tau}(t)}^{2}={\dot{\tau}_{0}}^{2}{h(t)\over{h(\tau_{*})}}
{g_{\tau\tau}(\tau_{*})\over{g_{\tau\tau}(t)}}}
{}From the profile of ${h\over{g_{\tau\tau}}}$ it follows that the brane
accelerates gradually toward the tip of the conifold.
For large $\tau$ we may use the KT radial coordinate $r$ \rdef, in terms
of
which \taudot\ is ${{d^{2}r}\over{d\xi^{2}}}=0$, which
is another expression of the balancing of gravitational forces and forces
due to flux.

\subsec{The Induced Cosmology}

An observer on the brane naturally sees an induced metric
\eqn\branemet{ds_{\rm{brane}}^{2} ~=~ -dt^2 + h^{-1/2}(\tau) (dx_1^2 +
dx_2^2
+ dx_3^2)~.}
But given that the brane trajectory is a function $\tau(t)$,
\branemet\ gives rise to a standard FRW cosmology
\eqn\FRW{ds^2 = -dt^2 + a^2(t) (dx_1^2 + dx_2^2 + dx_3^2)}
with $a(t)$ given by
\eqn\scalefac{a(t) = h^{-1/4}(\tau(t)) .}

Notice that the graviton wavefunction has a $\tau$-dependent
overlap with a brane located at various points in the metric
\ksmetric.  This is simply the effect exploited in \RS.
The dimensionless strength of gravity therefore scales according to
\eqn\generalg{G_N(t)m_{{open}}^{2} \sim h(\tau(t))^{-{1\over{2}}} \sim
a(t)^{2}} \noindent
where $m_{{open}}$ is the mass of the first oscillating open string
mode.  A physicist residing on the brane may choose to fix {\it{one}} of
the dimensionful quantities $G_N$, $m_{{open}}$ in order to set his units
of length.  Grinstein et al. \grinstein\ have shown that a brane observer
who uses proper distance to measure lengths on the brane will necessarily
find
fixed masses and variable $G_N$.  One can argue for the same system of
units by
stipulating that elementary particle masses
should be used to define the units, and should
be considered fixed with time.  
In this model we will use the mass of the first excited open string mode 
to fix 
such a frame; in a more realistic model, one would want other 
(perhaps ``standard model'') degrees of freedom to be 
the relevant massive modes.

A brane observer following an inward-falling trajectory in the background
\ksmetric\ would therefore make the following statements.

\noindent 1. Elementary particle masses, e.g. $m_{{open}}$,  are
considered fixed with time.

\noindent 2. In these units, the proper distance between galaxies on the
brane
scales with $a(t)$ as in standard FRW cosmology.  In consequence, for the
infalling
brane (moving towards $\tau=0$) one observes blueshifting of photons.

\noindent 3. The gravitational coupling on the brane is time-dependent,
\eqn\tgn{G_{\rm{N}}(t)\sim a(t)^{2}~.}
Therefore, as the universe collapses,
the strength of gravity decreases.

In fact, \scalefac\ together with \tgn\ imply that in 4d $\it Planck$
units, the size of the universe remains $\it fixed$.
{}From this ``closed string'' perspective, the cosmology is particularly
trivial; the brane radial position is described by a scalar field
$\Phi_{r}$ in the 4d action which is undergoing some slow time
variation (and, for small brane velocity,
carries little enough energy that backreaction is not
an issue).  However, in this frame particle masses vary with time.
We find it more natural, as in \grinstein, for a
brane observer to view physics in the frame specified by 1-3 above;
we will henceforth
adopt the viewpoint of such a hypothetical brane cosmologist.
In \S3.1 we describe the field redefinition which takes one from
the ``brane cosmologist'' frame to the ``closed string'' frame
in a toy model.

\medskip
\noindent{\it{The Bounce}}
\medskip

As the brane falls from $\tau^*$ towards zero, the scale factor
decreases monotonically.  It hits $\tau=0$ in finite proper time.
However, as is clear from the metric \ksmetric, there is no real
boundary of the space at this point; $\tau=0$ is analogous to the
origin in polar coordinates.  The brane smoothly continues
back to positive $\tau$, and the scale factor re-expands. Although
it is hard to provide an analytical expression for $a(t)$ given
the complexity of the expressions \his\ and \iis, we can
numerically solve for $a$; a plot appears in Figure 1.

\fig{The scale factor $a(t)$ as a function of proper time for a brane
near the tip of the Klebanov-Strassler geometry.  This particular bounce
begins
from radial position $\tau = 4$.  }{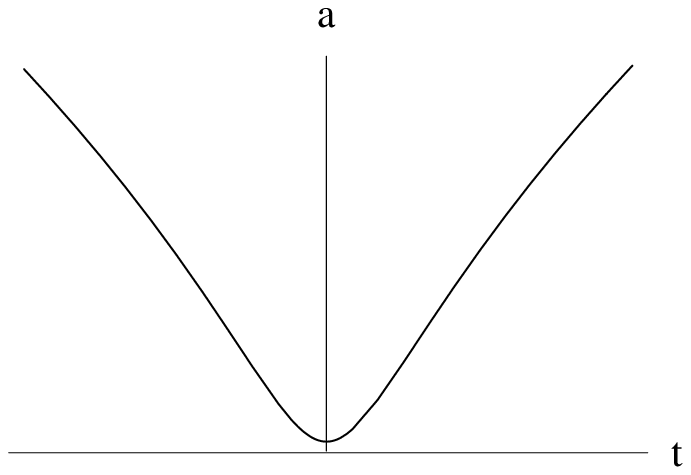}{3.3truein}

In the approximate supergravity dual to the cascading gauge theory
studied in \KT, there is instead a naked singularity in the
region of small $\tau$, which is deformed away by the fluxes
\flux.  In the KT approximation to the physics, then, the
cosmology on the brane would actually have a spacelike singularity
at some finite proper time. The evolution in this background
agrees with Figure 1 until one gets close to the tip of the
conifold; then, in the ``unphysical'' region of the KT solution,
the brane rapidly re-expands, and a singularity of the curvature
scalar of the induced metric arises at a finite proper time.  A
plot of $a(t)$ for this case appears in Figure 2.

\fig{The scale factor $a(t)$ as a function of proper time for a brane
near the singularity of the Klebanov-Tseytlin geometry.  The explosive
growth of $a(t)$ on the right coincides with a curvature singularity in the
induced metric.}{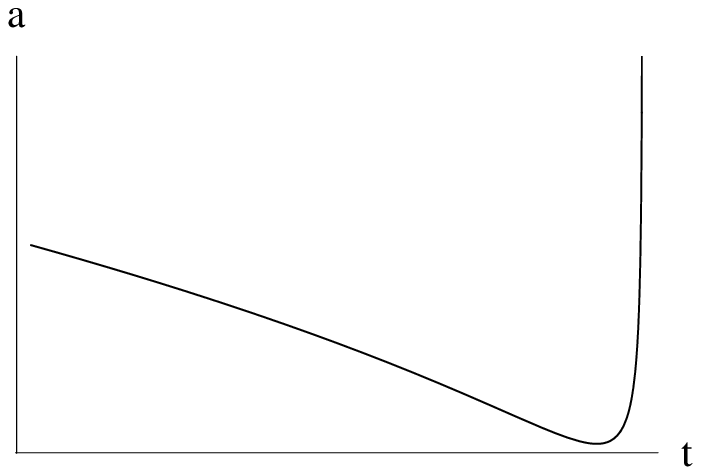}{3.3truein}

Hence, we see that string theory in the smooth KS background gives
rise to a bouncing brane cosmology, while the KT approximation
would have given rise to a cosmology with a spacelike crunch.
There has been great success in understanding the
resolution of timelike singularities in string theory, so it is
heartening to see that in some special cases one can translate those
results to learn about spacelike singularities as well.

\medskip
\noindent{\it{Limiting behaviors}}
\medskip

In the two asymptotic regimes of $\tau \sim 0$ and very large
$\tau$, the formulae simplify \KO\ and the behavior of $a(t)$ can
be given explicitly.  For small $\tau$, the geometry is just the product
\kstwo.
Hence, in this limit, the brane is effectively falling in an
${\it unwarped}$ 5d space, and the cosmology is very simple:
\eqn\branelim{a(t) = {\rm constant} + {\cal O}(t^2)~.}

In the large $\tau$ regime, the metric \otherlim\ differs from $AdS_5$
by logarithmic corrections, and so the brane trajectory deviates very
gradually from
that of a D3-brane in AdS.  For simplicity we present here the induced
cosmology on
a D3-brane in AdS; the logarithmic corrections require no new ideas but
lead to more
complicated formulae.
{}From \dbi, using the D3-brane form of the $AdS_5$ metric
\eqn\ads{ds^{2} = r^{2}(-d\xi^{2} + dx_1^2 + dx_2^2 +
dx_3^2)+{{dr^{2}}\over{r^2}}}
we find, in terms of proper time,
\eqn\adsa{a^{2}(t)=a^{2}(0)(1+2{\dot{r_{0}}\over{r_{0}}}t)}
for a brane with initial position and velocity $r_{0}$, $\dot{r_{0}}$ at
$t=0$.
It follows that
\eqn\adsac{ ({{\dot{a}}\over{a}})^2={C\over{a^{4}}}}
where $C={a^4(0){({{\dot{r_{0}}\over{r_{0}}}})}^{2}}$.
Because the right hand side of \adsac\ scales like the energy density of
radiation, this has been termed
``dark radiation'' \refs{\binetruy,\mukohyama}.  In the language of \KK\ 
it
might also be called ``mirage
matter with equation of state $\rho=3p$.''

The Friedmann equation \adsac\ has been thoroughly investigated in the
context of
Randall-Sundrum models.  In particular, just such a law was found to arise
on a
visible brane which is separated from a Planck brane by an interval whose
length varies
with time (see \khoury\ and references therein).  This is entirely
consistent with our scenario,
as the Calabi-Yau provides an effective Planck brane and the bulk motion
of the probe changes the length of the
interval between the branes.

As the brane proceeds to larger $\tau$, eventually it will reach the
region where the KS throat has been glued onto a Calabi-Yau space.
Beyond that point it is no longer possible for us to say anything
universal
about the behavior of the brane cosmology.

\subsec{Issues of backreaction}

There are several issues involving backreaction that merit consideration. 
To argue that the bounce we have seen in \S2.4\
accurately describes the behavior of the brane as it propagates in
from $\tau_*$ and back out again, we must ensure that
the state with nonzero $\dot{\tau}$ on the brane does not contain
enough energy to significantly distort the closed string
background geometry.  
In fact we must check both that a motionless 
brane in the throat creates a negligible backreaction, and that 
the kinetic energy on the brane does not undergo gravitational
collapse (yielding a clumpy brane) on the
relevant timescales.   
It is also important to understand the extent of
the backreaction from
semiclassical particle production.
Finally, the presence of nonzero energy density on the brane leads to
a potential for the Calabi-Yau volume modulus (as in \S6\ of \anti). 
We will imagine that this modulus has
been fixed and will neglect this effect. 

The first concern can be dismissed quickly. In the limit of small $g_s$
the backreaction on the
closed string
background is small.
The second concern needs to be discussed in somewhat more detail.
The falling brane necessarily has energy
density localized on its worldvolume.
After a sufficiently long time this initially uniform energy can become
inhomogeneous because of the Jeans instability.
In this subsection we demonstrate that, for a suitable choice of the
parameters of the KS geometry, this instability is
negligible during the bounce portion of the history of the
brane universe. 

\medskip
\noindent{\it{Jeans Instability}}
\medskip

For a uniform fluid of density $\rho$, the Jeans instability appears at
length scales greater than
$L_{Jeans}\equiv {v_{s}\over{\sqrt{\rho G_{N}}}}$, where $v_{s}$ is the
velocity of sound.
Perturbations with this wavelength could destabilize the brane given a
time $t_{instability}\geq{L_{Jeans}}$.
In terms of the volume $V_{6}$ of the Calabi-Yau,
\eqn\gneff{ G_{N}=g_s^{2} l_s^{8}{V_{6}}^{-1} h(\tau_{UV})^{{1\over{2}}}
h(\tau)^{-{1\over{2}}}}
where we choose $\tau_{UV}$ such that $r_{UV}$ (as given in \rdef)
is of order one (so the throat extends slightly into the KT regime before
gluing into the Calabi-Yau).
For the compactifications of interest $V_{6}\geq l_s^{6}$,\foot{
In fact, as discussed in \GKP, warped compactifications really
reproduce the RS scenario when the volume is not very large in string
units (since the flux and brane backreaction which
produce the warping
become larger effects at small Calabi-Yau volume). We are
assuming we are at the threshold volume where the warping
becomes a significant effect, which should justify the estimate \gneff.}
so that for $\tau
\leq \tau_{UV}$
\eqn\gbound{ G_{N}\leq {g_{s}}^{2} {l_{s}}^{2}~.}
{}From \dbi, \taudot, we see that the energy density on the brane is
constant,
\eqn\rhobrane{ \rho={1\over{2 g_s^{2}
l_s^{4}}}{h(\tau_{*})^{-1}g_{\tau\tau}(\tau_{*}){\dot{\tau_{0}}^{2}}}}
so
\eqn\tjeans{ t_{instability} \geq
{{v_{s}\over{\dot{\tau_{0}}}}h(\tau_{*})^{{1\over{2}}}
g_{\tau\tau}(\tau_{*})^{{-{1\over{2}}}}}l_{s}~.}
Because the brane accelerates toward the tip of the conifold, to fall from
$d_{*}$ to the
tip and rebound requires a time
\eqn\tbounce{ t_{bounce}\leq {{2d_{*}}\over{v_{0}}}~.}
This leads to (we now drop numerical factors of order one)
\eqn\aratio{ {{t_{bounce}}\over{t_{instability}}} <
{{d_{*}}\over{l_{s}}} h(\tau_{*})^{-{1\over{2}}}~.}
Using the asymptotic form of $I(\tau)$, $K(\tau)$ we find
\eqn\bratio{ {{t_{bounce}}\over{t_{instability}}} <
{1\over{\sqrt{g_{s}
M}}}{\tau_{*}}^{3\over{4}}
{l_{s}}^{-2}{({\epsilon^{2}e^{\tau_{*}}})}^{2\over{3}}~.}

Because we have glued the KS throat into the Calabi-Yau geometry
at a location where $r=r_{UV}$ of \rdef\ is of order one, we see
that $\epsilon^{2}e^{\tau_*} = {\cal O}(1)$.   
This leads to
\eqn\bratiot{{{t_{bounce}}\over {t_{instability}}} <
{1\over {\sqrt{g_s M}}}\tau_{*}^{3/4}~.}
Finally, since the hierarchy between the UV and IR ends of the throat is
exponential in $\tau_*$, it is natural to take $\tau_*$
to be a number of order 5-10 (in the language of RS scenarios, $\tau_*$
controls the length of the interval in AdS radii, up to factors of
$\pi$).
Therefore, in the supergravity regime where $g_s M >> 1$, \bratiot\
demonstrates that we can neglect the Jeans instability on the brane in
discussing the dynamics during the bounce.

\medskip
\noindent{\it{Particle Creation}}
\medskip

Because the bounce cosmology is strongly time-dependent, it is 
also important
to
consider the spectrum of particles created semiclassically by the bounce.
We
will argue that the energy density due to such particle production is
small enough
that its backreaction is negligible.

The bounce geometry \FRW\ is conformally trivial, so massless, conformally
coupled
scalar fields will not be produced by the cosmological evolution.  Massive
fields
break the conformal invariance.  The relevant massive
scalar fields
on the brane are excited string states with mass $m \ge
{1\over{l_{s}}}$. 
Quite 
generally
we expect that modes with frequencies $\omega \gg {\dot{a}\over{a}} \equiv
H$ will not be significantly
populated by the bounce, i.e. the probability that a comoving detector
will register such
a particle long after the bounce is exponentially small in
${\omega\over{H}}$.  The cases of
interest involve slow-moving branes, so the maximum value of $H$ is far
below the string scale.
Thus we expect the energy density due to particle creation should be
quite small.

Concrete calculations of the production of massive scalar and fermion
fields in a bouncing $k=0$ FRW cosmology were carried out in \audschaf\
(though the system in consideration there did not satisfy Einstein's
equations).  The scale factor in \audschaf\ has the same
limiting behaviors as our own, and the results there
are consistent with our expectations.
It would be interesting to carry out the relevant particle creation 
calculation
directly in string theory.  A particle creation calculation
in closed string theory was described in worldsheet (2d conformal
field theory) language in \evaetal.

\newsec{Four-dimensional Lagrangian description}

\subsec{Effective Lagrangian}
In the limit of low matter density on the probe brane, the
cosmology is determined entirely by the bulk geometry.  The
D3-brane trajectory is determined by the Born-Infeld action, and
the induced metric along this trajectory provides a time-dependent
mirage cosmology.  The mirage cosmology proposal of \KK\ includes another
step: one can write down the Friedmann equations for the cosmology
and identify the right hand side with mirage density and mirage
pressure.

This is not yet an ideal formulation from the perspective of a brane
resident.  One would like a four-dimensional Lagrangian description
of the mirage matter, of the cosmological evolution, and of the
variation of $G_N$.  In particular, since a bounce in a flat
Friedmann-Robertson-Walker universe necessitates violation of the
null energy condition, it would be interesting to understand
this violation in terms of a 4d Lagrangian and energy-momentum
tensor.  In this section we will 
propose a toy scalar-tensor Lagrangian which admits
cosmologies reproducing the basic features of
our ``bouncing brane'' solutions; similar Lagrangians have arisen
in the study of RS cosmology \jatkar.

The massless fields in our 4d theory include a 4d graviton and
the massless open strings on the D3-brane: a U(1) gauge field
$A_{\mu}$, a scalar $\Phi_{r}$
corresponding to radial motion in the compactified throat, and 
scalars $\Phi_{i}$, $i = 1,\cdots,5$ 
parametrizing motion in the angular coordinates.  All other scalar fields
are massive. (In fact without a no-force condition there can be
a potential and a mass for $\Phi_{r}$. For simplicity we will work
only with the BPS case, but the trajectory of anti-branes in the KS
throat would also yield an interesting time-dependent solution.\foot{In 
particular, anti-branes
near the tip of the conifold can annihilate by merging with flux \anti. 
This could potentially
lead to a cosmology which begins or ends with a tunneling or annihilation 
process.})  We
will choose to fix the $\Phi_{i}$,
and the requirement of negligible energy density in open string
modes on the brane means that $A_{\mu}$ is not relevant for
cosmological purposes.  This leaves $\Phi_{r}$ and $g_{\mu\nu}$ as the 
only massless
fields entering the 4d Lagrangian.

Our goal in this section is to show explicitly how an observer who sees 
particle masses which depend on $\Phi_{r}$ could
change his units of length and see an FRW cosmology with varying $G_N$.  
(In \S2.4 we provided several arguments motivating this choice of frame.)
Because the full Lagrangian for a brane observer in the KS background, 
including all massive fields, is quite complicated,
it will be most practical to work with a simpler Lagrangian which has the 
correct schematic features.  In particular, all particle
masses depend on $\Phi_{r}$ in the same way, so it will suffice to 
consider a single massive field $\chi$
(which could be, for example, an excited open string mode).

A ``mass-varying'' Lagrangian with the appropriate features is
\eqn\mvaction{L = \int d^3 x \sqrt{-g}~({R\over{16 \pi G_N}} 
-{R\over{12}}{\Phi_{r}^{2}}-{1\over{2}}g^{\mu\nu}\nabla_\mu \Phi_{r} 
\nabla_\nu \Phi_{r}-
{{1\over{2}}g^{\mu\nu}\nabla_\mu \chi \nabla_\nu \chi - 
{1\over{2}}m^{2}(\Phi_{r})\chi^{2}-V(\chi)})}
where $\chi$ is a matter field on the brane whose mass depends on 
$\Phi_{r}$ as
\eqn\massofchi{m^{2}(\Phi_{r}) \equiv{\Omega^{2}(\Phi_{r})}\mu^{2}}
for fixed $\mu$.
The form of the potential for $\chi$ and the coupling of $\chi$ to the 
curvature scalar will be unimportant for this analysis, and we will 
henceforth
omit these terms.  Note that $\Phi_{r}$ is conformally coupled. 

As discussed in \S2.4, an observer confined to the brane most naturally 
holds fixed the masses of fields on the brane.
This can be accomplished by performing the change of variables
\eqn\changeg{\tilde{g}_{\mu\nu}=\Omega^{2}(\Phi_{r}){{g}_{\mu\nu}}}
\eqn\changep{\tilde{\Phi}_{r}=\Omega^{-1}(\Phi_{r}){\Phi_{r}}}
\eqn\changec{\tilde{\chi}=\Omega^{-1}(\Phi_{r}){\chi}}

\noindent The resulting ``mass-fixed'' Lagrangian is
\eqn\mfaction{
\eqalign{
L = \int d^3 x \sqrt{-\tilde{g}}~(&{\tilde{R}\over{16 \pi G_N 
\Omega^{2}(\Phi_{r})}} 
+ {3\over{8\pi G_{N} \Omega(\Phi_{r})^4}} \tilde g^{\mu\nu}\nabla_{\mu}
\Omega \nabla_{\nu}\Omega -  
{\tilde{R}\over{12}}{\tilde{\Phi}_{r}^{2}}\cr 
&-{1\over{2}}\tilde{g}^{\mu\nu}\nabla_\mu \tilde{\Phi}_{r} \nabla_\nu 
\tilde{\Phi}_{r}-
{1\over{2}}\tilde{g}^{\mu\nu}\nabla_\mu \tilde{\chi} \nabla_\nu 
\tilde{\chi} - {1\over{2}}{\mu}^{2}\tilde{\chi}^{2})~.}}
We have discarded terms which look like $(\nabla\Omega)^{2}\tilde{\chi}^{2}$
because $\dot{\Omega} << \mu$ (at least in our example, where $\chi$
represents a massive string 
mode).
Terms which look like $(\nabla{\Omega})^2 \tilde \Phi_{r}^2$ 
cancel due to
the conformal coupling of $\Phi_{r}$.

The effective gravitational coupling is given by 
\eqn\geff{ {G_N}^{eff} =  
G_N \Omega^{2}(\Phi_{r})~.    }
According to the discussion in \S2.4, we expect that $\Omega^{2}(\Phi_{r}) 
= h(\tau(\Phi_{r}))^{-{1\over{2}}}$, so indeed the
strength of gravity scales as required by \tgn.  (We will not need the 
explicit relation between $\tau$ and $\Phi_{r}$.)

We are interested in the limit where the backreaction due to 
$\tilde{\Phi}_{r},\tilde{\chi}$ is small, so in particular
$\tilde{\Phi}_{r},\tilde{\chi} \ll m_{Planck}^{eff}$.  This means that for 
the purpose of solving the Einstein equations
in the mass-fixed frame we may neglect terms which are suppressed by a 
factor of ${G_N}$.
Defining 
\eqn\gamdef{\gamma=\sqrt{{3\over{4 \pi G_N }}}\Omega^{-1}(\Phi_{r})} 
we may 
write the effective Lagrangian

\eqn\effaction{L = \int d^3 x 
\sqrt{-\tilde{g}} \left({\tilde{R}\over{12}}{\gamma^2}+
{1\over{2}}\tilde{g}^{\mu\nu}\nabla_\mu 
\gamma \nabla_\nu \gamma
+{\cal{O}}({{\Phi}_{r}\over{m_{Planck}}})~\right).}
Observe that the kinetic energy term is now negative semidefinite (we are 
using signature $-+++$), so it is easy to violate the null energy condition
which is relevant (via the singularity theorems) 
in constraining the behavior of the 
metric $\tilde g_{\mu\nu}$.\foot{Notice that because of the non-minimally
coupled scalar, it is also possible to violate the null energy condition 
which governs the behavior of $g_{\mu\nu}$.}

The equation of motion which follows from this Lagrangian is
\eqn\rhoeom{\tilde{g}^{\mu\nu}\nabla_{\mu}\nabla_{\nu}{\gamma} - 
{\tilde{R}\over{6}}\gamma={\cal{O}}({{\Phi}_{r}\over{m_{Planck}}})}
Now let us see that this system reproduces our expectations from
\S2.4.  
Given an FRW cosmology specified by $a(t)$, if we set
$\gamma(t)={c {a^{-1}(t)}}$ for some 
constant $c$
then \rhoeom\ is satisfied identically.  {}From \changeg, \changep, \changec\
it is clear that we should identify 
\eqn\cosmo{a(t) \propto \Omega({\Phi_{r}(t)}).}
Then the Einstein equations for \effaction\ are satisfied if 
the varying-mass metric $g_{\mu\nu}=\eta_{\mu\nu}$ and the
mass-fixed metric $\tilde{g}_{\mu\nu}=a^{2}(t)\eta_{\mu\nu}$.  
So as 
discussed in \S2.4, we have two complementary perspectives: the brane 
observer uses
the mass-fixed action \mfaction\ and sees an FRW cosmology with varying 
$G_{N}$,
while the ``closed string'' observer sees gravity of fixed strength in 
Minkowski space.

\subsec{Relation to Warped Backgrounds}

We can be slightly more explicit about how the toy model of \S3.1\ would
be related to a given warped background.
Given any function $a(t)$, we can construct a warped background $h(r)$ 
such
that a no-force brane probe of that geometry experiences an induced
cosmology specified by $a(t)$.
We simply define $\xi=\int{dt\over{a(t)}}$, $r=v\xi$ ($v$ constant), and
$h(r)={a(r)}^{-4}$.

A few comments are in order: 

\noindent 1. Very few backgrounds $h(r)$ will correspond to solutions of 
IIB
supergravity.  One which does, and indeed corresponds to a D3-brane in the
warped deformed conifold, is given by taking $\tau(t)$ to solve \taudot\
and setting ${a(t)}={h^{-{1\over{4}}}}(\tau(t))$ with $h$ given by \his.

\noindent 2. The no-force condition is only a convenience.  We could 
instead take
$r(\xi)$ to be any function of $\xi$.  This would correspond to a brane
which accelerates due to external forces.  Again, very few systems of this
sort arise from known branes of string theory moving in
valid supergravity backgrounds.

\newsec{Discussion}

As demonstrated in general terms in \S3, and in a special example in string
theory in \S2, in the presence of scalar fields 
it is easy to evade the singularity
theorems (from the perspective of a reasonable class of observers), even 
with
a $k=0$ FRW universe.
It therefore seems likely that many examples of such constructions,
arising both as cosmologies on D-branes and perhaps even
as closed string cosmologies,
should be possible.  The cosmology we presented is just a slice of
evolution between some initial time when we join the brane
moving down the throat, and a final time when it is heading into the
Calabi-Yau region.  The later evolution of our model is then
non-universal; it depends on the details of the Calabi-Yau model
(or in the language of \RS, the detailed structure of the Planck brane).
It would be very interesting to write down models
with 4d gravity whose dynamics can be controlled for an eternity;
some controlled, eternal closed string cosmologies were recently described
in \evaetal.

The cosmology discussed here is far from realistic.
As a first improvement, one would like to study probe branes with a spectrum
of massive fields below the scale ${1\over {l_{s}}}$ (which could be 
called ``standard model'' fields).  It may be possible to construct
such examples by using parallel D3-branes which are slightly separated in the
radial direction, wrapped Dp-branes with ${\rm p}>3$, 
or 
anti-branes in appropriate regimes.   
It is also important to control the time-variation of
$G_N$ during/after nucleosynthesis, since this is highly constrained by
experiment 
(see for instance \globular).
To improve the situation, one can envision a
program of ``cosmological engineering.''  That is, one could 
try to design IIB solutions with
background fields specifically chosen to give rise to interesting mirage
cosmologies (various authors have already proposed mirage models
of closed universes \youm, inflation with graceful exit \typez,
asymptotically de Sitter spaces \medved, etc., though most of these
models do not include 4d gravity).
Each desired feature of the cosmology would result in a new condition
on the closed string fields.  Then one would simply 
impose these conditions along with the field equations of IIB
supergravity.

\centerline{\bf{Acknowledgements}}

We are indebted to L. Susskind for lengthy discussions and for asking
several important questions.
We would also like to thank O. Aharony, M. Fabinger, S. Giddings, G. 
Horowitz,
N. Kaloper, E. Silverstein and
H. Verlinde for helpful discussions.   We are also grateful to C. Herzog for
pointing out a minor error in the first version of this paper.
This work was
supported in part by the DOE under contract DE-AC03-76SF00515. The
work of S.K. is also supported by a David and Lucile Packard
Foundation Fellowship for Science and Engineering, an Alfred P. Sloan
Foundation Fellowship, and National Science Foundation grant PHY-0097915.
The
work of L.M. is supported in part by a National Science Foundation
Graduate Research Fellowship.

\listrefs

\end